\documentclass[12pt]{article}
\usepackage[dvips]{graphicx}
\newcommand{\krzlnbrb}{\ensuremath{{\Gamma(\krzlndk{}) \over \Gamma(\kpipi{})} }}
\newcommand{\kpimndk}{\ensuremath{D^+ \rightarrow K^- \pi^+ \mu^+ \nu }}
\newcommand{\krzb}{\ensuremath{\overline{K}^{*0}}}
\newcommand{\krz}{\ensuremath{K^{*0}}}
\newcommand{\krzlndk}{\ensuremath{D^+ \rightarrow \krzb \ell^+ \nu_\ell}}
\newcommand{\krzmndk}{\ensuremath{D^+ \rightarrow \krzb \mu^+ \nu}}
\newcommand{\kpipi}{\ensuremath{D^+ \rightarrow K^- \pi^+ \pi^+ }}
\newcommand{\gevcsq}{\ensuremath{\textrm{GeV}/c^2}}
\newcommand{\thv}{\ensuremath{\theta_\textrm{v}}}
\newcommand{\thl}{\ensuremath{\theta_\ell}}
\newcommand{\costhv}{\ensuremath{\cos\thv}}

\newcommand{\costhl}{\ensuremath{\cos\thl}}

\newcommand{\qsq}{\ensuremath{q^2}}
\newcommand{\bw}{\ensuremath{\textrm{B}_{\krz}}}
\newcommand{\mkpi}{\ensuremath{m_{K\pi}}}

\def\beq{\begin{equation}}
\def\eeq#1{\label{#1}\end{equation}}
\def\eeqn{\end{equation}}

\def\beqa{\begin{eqnarray}}
\def\eeqa#1{\label{#1}\end{eqnarray}}
\def\eeqan{\end{eqnarray}}

\let\bar=\overbar

\def\etal{{\it et al.}}

\def\Dslash{\not{\hbox{\kern-4pt $D$}}}
\def\dslash{\not{\hbox{\kern-2pt $\del$}}}

\def\msb{{\bar{\ssstyle M \kern -1pt S}}}

\def\Title#1{\begin{center} {\Large {\bf #1} } \end{center}}

\begin{document}

\Title{Charm Semileptonic Decays}

\bigskip\bigskip


\begin{raggedright}  

{\it Jim Wiss\index{Wiss J.}\\
Department of Physics\\
University of Illinois \\
Urbana, IL,61801, USA}
\bigskip\bigskip
\end{raggedright}

\begin{figure}[tbph!]
 \begin{center}
  \includegraphics[width=2.5in]{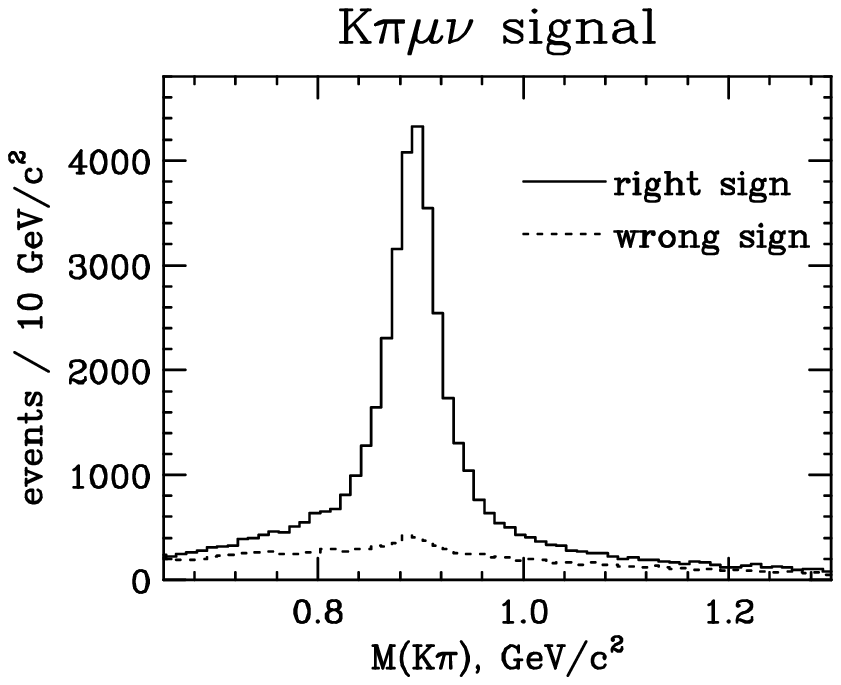}
  \includegraphics[width=2.5in]{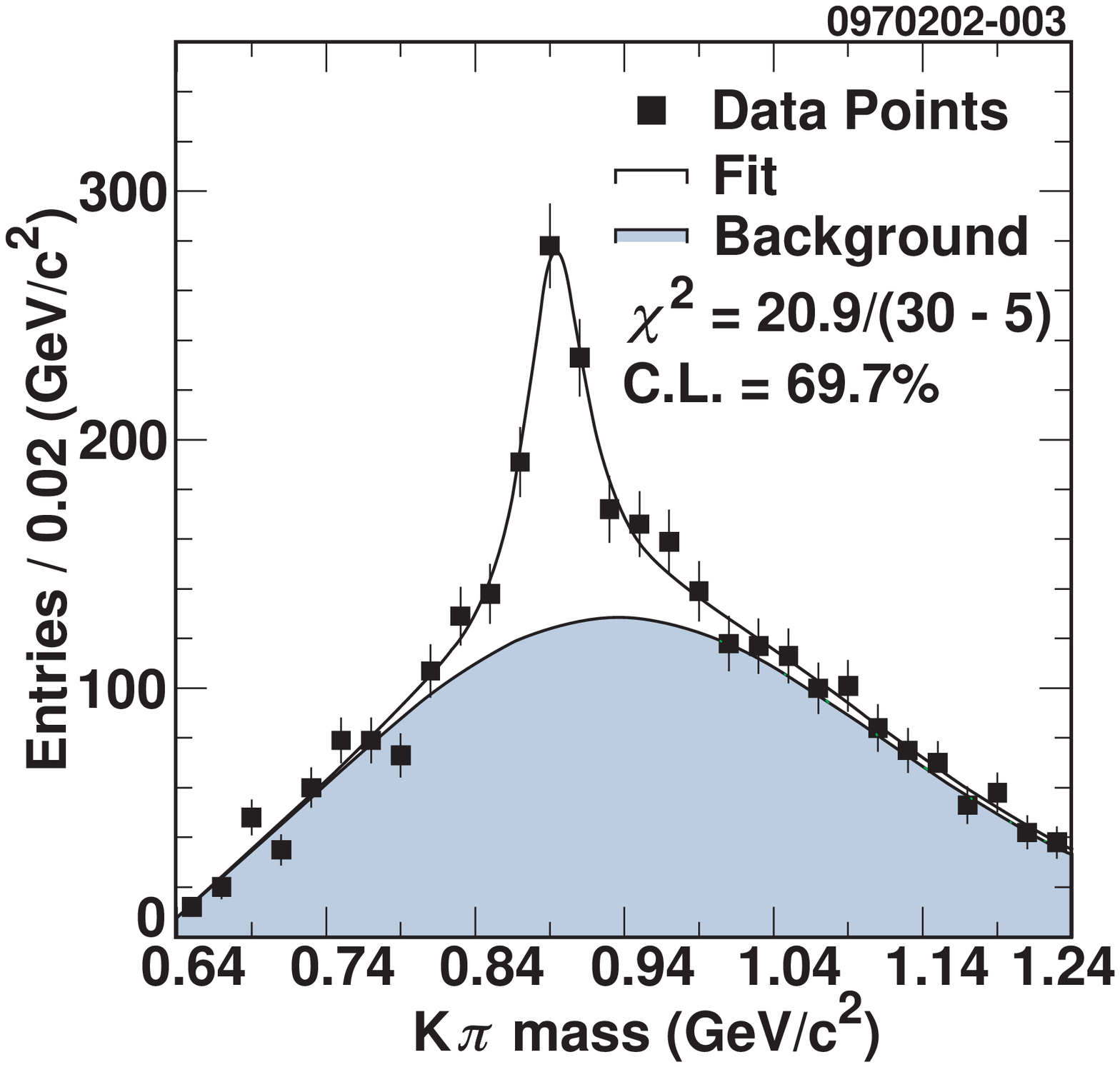}
 
  \caption{\kpimndk{} signal.  (left) The FOCUS right-sign and
  wrong-sign samples are shown. The wrong-sign-subtracted yield is
  $31\,254$ events. (right) A partial sample of $D^{*+} \rightarrow
  \pi^0 D^+ \rightarrow \pi^0 (\krzb e^+ \nu)$ from CLEO. This is the
  sample form one of their bins in the $D^* -D$ mass difference.
\label{signals}} \end{center}
\end{figure}

I discuss new data on charm semileptonic decay concentrating on two
topics involving the decay \kpimndk{}.  The first topic is the
observation of interference in this decay by the FOCUS
collaboration\cite{anomaly}.  The second are new measurements of
branching ratio of \krzlndk{} relative to \kpipi{} from
CLEO\cite{cleo} and FOCUS.  Fig. \ref{signals} shows the \krzlndk{} signals of
these two groups.

\section{Interference in \kpimndk{}}
\begin{figure}[tbph!]
 \begin{center}
  \includegraphics[width=2.0in]{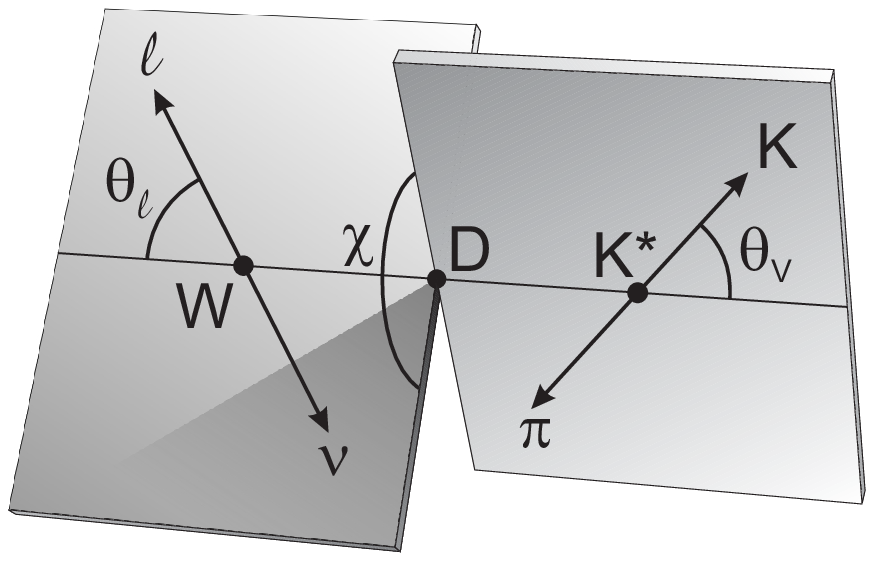}
 \includegraphics[width=3.5in]{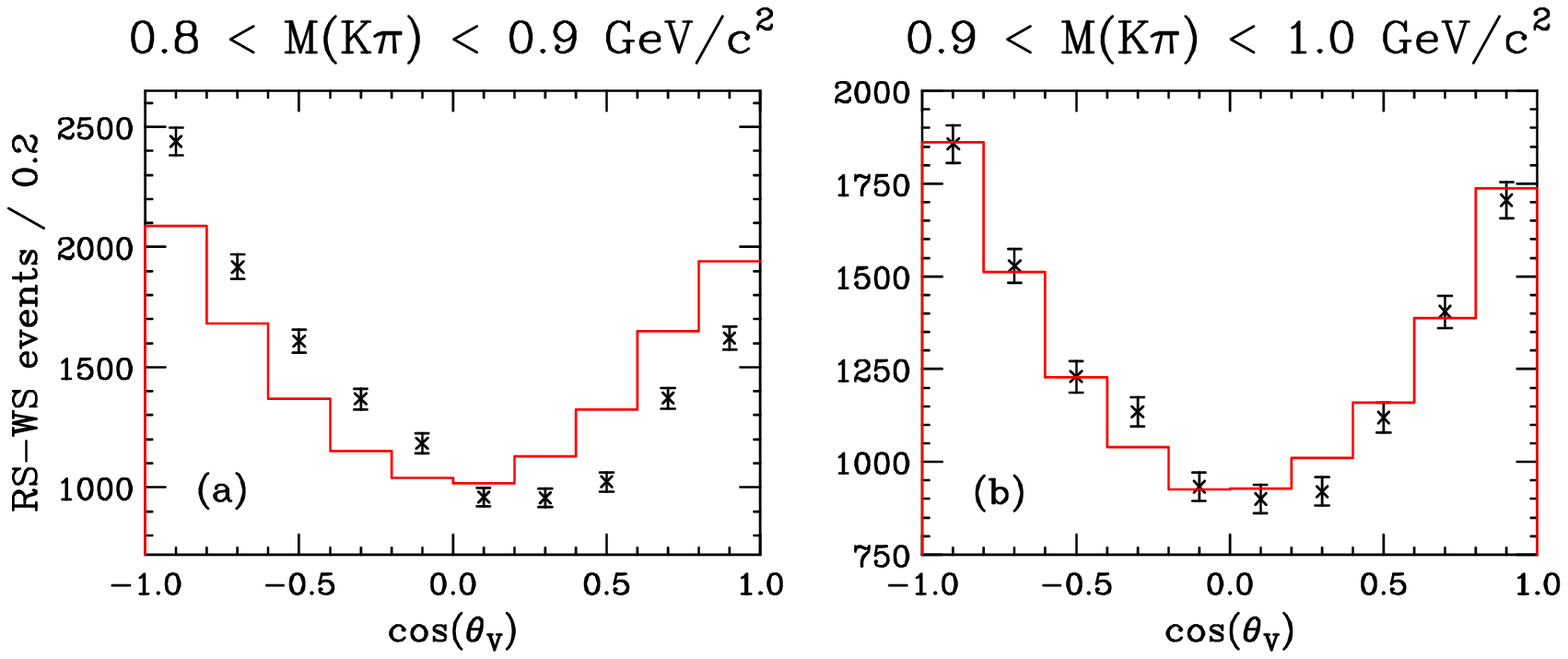} \caption{ (left)
 Definition of the three decay angles: \costhv{} is the decay angle of
 the kaon in the \krzb{} frame, \costhl{} is the angle of the charged
 lepton in the virtual $W^+$ frame.  $\chi$ is the acoplanarity angle
 between the two decay planes.  (right) Event distribution in
 \costhv{}, split between samples below and above 0.9 \gevcsq{}.  The
 points with error bars are (wrong-sign subtracted) FOCUS data and the
 solid histogram is a Monte Carlo simulation with efficiency and known
 charm backgrounds but no interference.
\label{angles}} \end{center}
\end{figure}

In our attempts to fit for the form factors controlling the decay
\krzmndk{}, we discovered a large, unexpected asymmetry in the
\costhv{} distribution shown in Fig. \ref{angles}. This asymmetry was
very strong for events with a \mkpi{} mass below the pole and weak for
events above the pole.
It was possible to understand the forward-backward asymmetry in
\costhv{} using the simple model summarized by Eqn.~\ref{amp1}.  Using
the notation of \cite{KS}, we write the decay distribution (in the
zero charged lepton mass limit) for \kpimndk{} in terms of the three
helicity basis form factors: $H_+~,~H_0~,~H_-$.  We have taken the
standard amplitude and added an interfering s-wave amplitude with a
constant modulus and phase ($A~\exp(i \delta)$) that interferes with
the \krzb{} Breit-Wigner (\bw{}) in the one place allowed by angular
momentum conservation.

\begin{equation}
{d^5 \Gamma \over dm_{K \pi}~d\qsq~d\cos\thv~d\cos\thl~d\chi}
\propto \qsq \left| \begin{array}{l}
 (1 + \cos \thl )\sin \thv e^{i\chi } \bw H_ +   \\
  - \,(1 - \cos \thl )\sin \thv e^{-i\chi } \bw H_ -   \\
  - \,2\sin \thl (\cos \thv \bw + Ae^{i\delta } )H_0  \\
 \end{array} \right|^2
\label{amp1}
\end{equation}
Assuming the s-wave amplitude is small (or the effect would have been
discovered already) it will be primarily observable through three
interference terms: 
\newline
$ 8\cos\thv\sin^2\thl A \Re\left(e^{-i\delta}\bw\right)H_0^2$ ,
  $-4(1+\cos\thl)\sin\thl\sin\thv\ A
  \Re\left(e^{i(\chi-\delta)}\bw\right)H_+H_0$, and
  $+4(1-\cos\thl)\sin\thl\sin\thv\ A
  \Re\left(e^{-i(\chi+\delta)}\bw\right)H_-H_0$.  Only the first of
  these terms will survive averaging over the acoplanarity, $\chi$.
  This was the term responsible for creating the \costhv{} asymmetry
  shown in Fig. \ref{angles} since it is proportional to \costhv{}.
  If we further weight our wrong-sign subtracted, azimuthally averaged
  data by \costhv{} , this is the only term that will survive
  in the full decay amplitude (given our nearly uniform
  angular acceptance).  It will have a distinct dependence on the
  \mkpi{} mass: $\Re\left(e^{-i\delta}\bw\right)$, as well as on
  \costhl{}: ($1 - \cos^2{\thl}$).

\begin{figure}[tbph!]
 \begin{center}
  \includegraphics[width=2.5in]{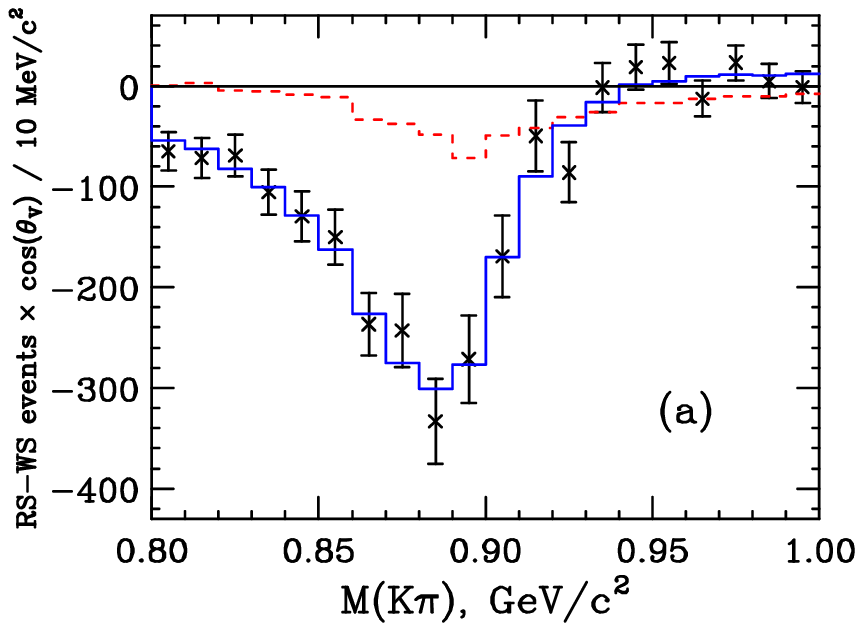}
  \includegraphics[width=2.5in]{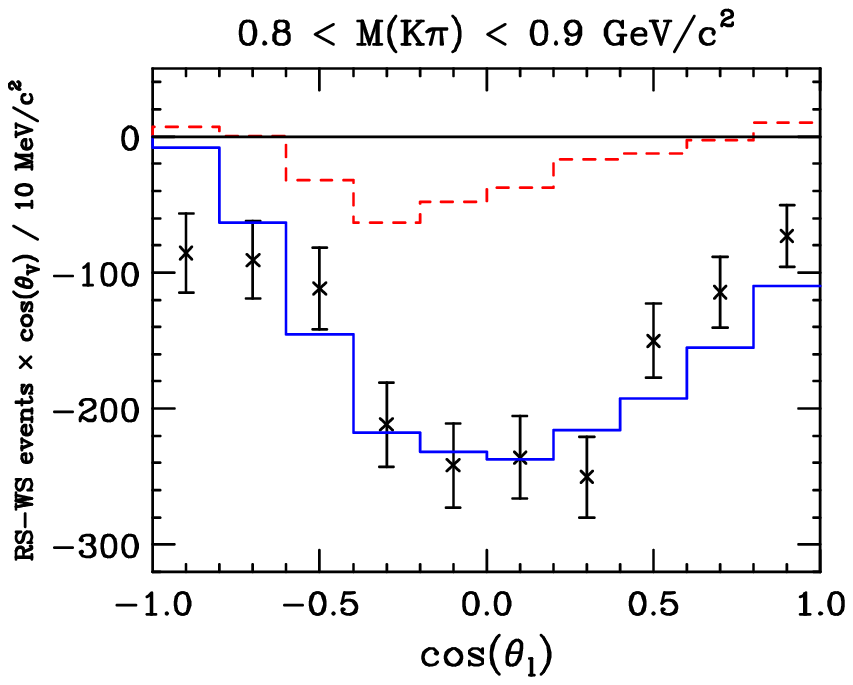}
  \caption{(left) The \costhv{}-weighted distribution in $K\pi$
  invariant mass. The data are the points with error bars. The dashed
  (red) histogram have no s-wave amplitude (null hypothesis). The solid (blue)
  histogram includes an s-wave amplitude of $0.36~\exp(i\pi/4)~GeV^{-1}$.
  (right) The \costhv{}-weighted distribution in \costhv{} compared to
  the null (red) and amplitude (blue) simulations.  } \label{asym}
  \end{center}
\end{figure}

Figure \ref{asym} shows two \costhv{}-weighted, wrong sign subtracted
distributions for $K \pi \mu \nu$. The left plot is the asymmetry
weighted \mkpi{} distribution which should resemble
$\Re\left(e^{-i\delta}\bw\right)$. For $\delta = 0$,
$\Re\left(e^{-i\delta}\bw\right)$ is odd function of \mkpi{} -
$m_{\krz}$ , while for $\delta = \pi/2$ this form is even in
\mkpi{} - $m_{\krz}$.  The data strongly resembles the expected plot
for $\delta = \pi/4$.  The right half of Fig. \ref{asym} is the
asymmetry weighted \costhl{} distribution with masses in the region
$0.8 < \mkpi < 1.0~\gevcsq{}$. It resembles the expected parabola in \costhl{}
with some modulation due to acceptance and resolution.

\begin{figure}[tbph!]
 \begin{center} \includegraphics[width=3.7in]{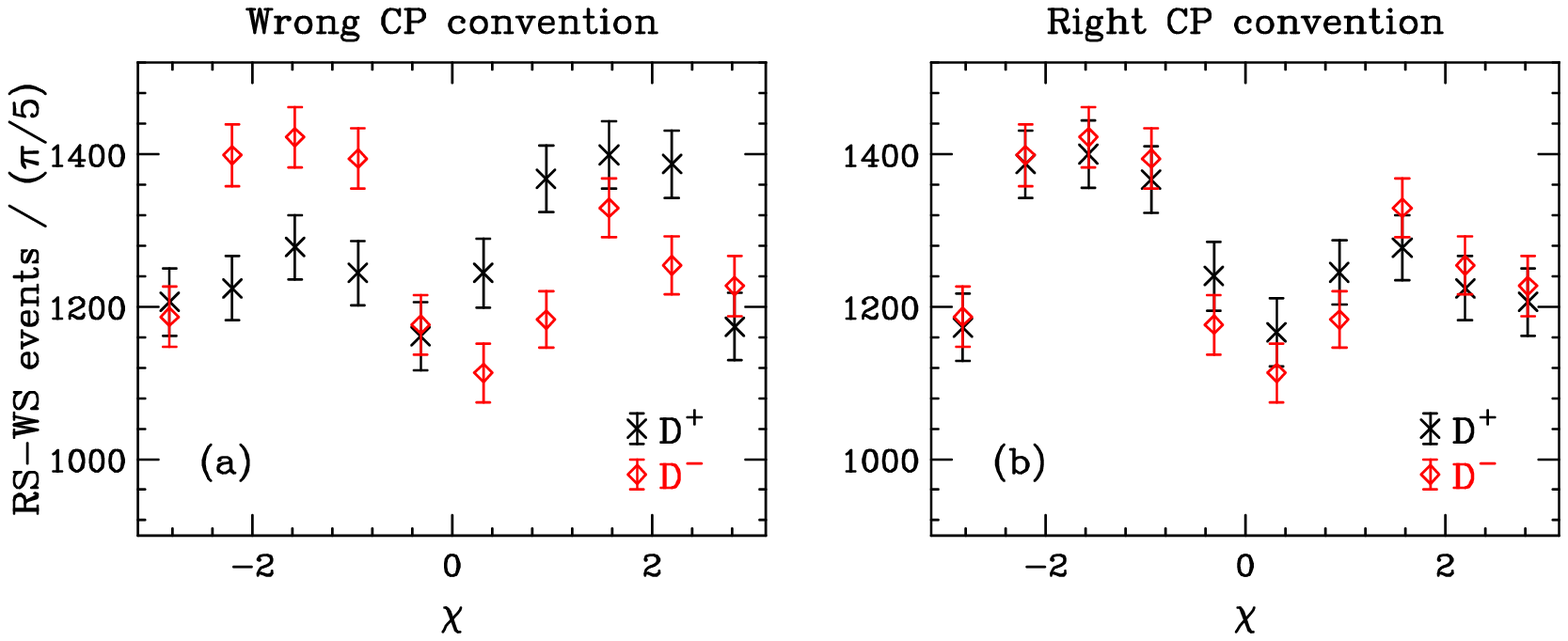}
\includegraphics[width=1.8in]{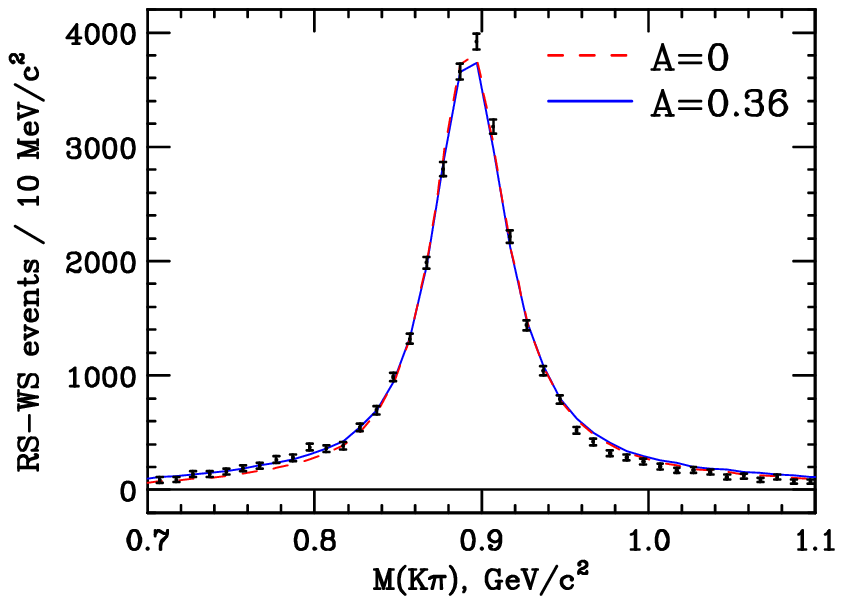} \caption{(left) The
wrong-sign subtracted acoplanarity distribution separated by
charm. The ``x'' points are for the $D^+$ while the ``diamond'' points
are for the $D^-$. (a) compares the distributions without the required
change in the $\chi$ convention as discussed above.  (b) compares the
distributions with the correct $\chi$ sign convention change.  (right)
The \mkpi{} mass distribution in data (with error bars) compared to
our null hypothesis (red) and s-wave (blue) Monte Carlos. The two
predicted \mkpi{} spectra are nearly identical.
\label{chidemo}} \end{center}
\end{figure}

In the absence of the s-wave interference, all acoplanarity dependent
terms in the \krzmndk{} decay intensity are functions of $\cos{\chi}$
and $\cos{2\chi}$.  The s-wave interference includes additional
acoplanarity dependent s-wave terms of the form: \newline
$+4(1-\cos\thl)\sin\thl\sin\thv\ A
\Re\left(e^{-i(\chi+\delta)}\bw\right)H_-H_0$ which brings 
in a $\sin{\chi}$ dependence thereby breaking $\chi
\leftrightarrow -\chi$ symmetry.  Figure \ref{chidemo} shows the
wrong-sign subtracted $\chi$ distribution separately for $D^+$ and
$D^-$ events in the range $0.8 < \mkpi{} < 1.0~\gevcsq{}$
Initially we were surprised by the inconsistency between the $D^+$ 
and $D^-$ acoplanarity until we realized that there is a sign
change in the $\chi$ convention between the particle and antiparticle.
After applying the correct convention, the $D^+$ and $D^-$ distributions
become consistent, and the odd $\chi$ contributions brought in 
through the s-wave interference
become very evident.
Why has the s-wave interference
in \krzlndk{} never been reported before, given that it has been
a process studied for nearly twenty years by several experiments?
One answer is that an amplitude of this strength and form
creates a very minor modulation to the \mkpi{} spectrum as shown
in Figure~\ref{chidemo}.  Another reason is that this effect is much
more evident when one divides the data above and below the \krzb pole.
Finally, the FOCUS
data set has significantly more clean \kpimndk{} events than
previously published data.
\section{New Measurements of \krzlnbrb{}}
\begin{figure}[tbph!]
 \begin{center} 
\includegraphics[width=4.in]{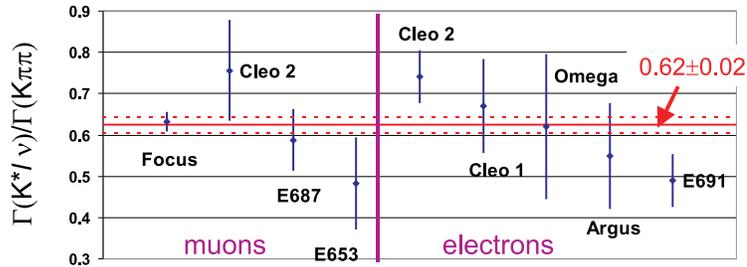}
  \caption{Summary of measurements on \krzlnbrb{}. The muon data, on the 
left, has been scaled by a factor of 1.05 to compare to the electron data. Our
preliminary FOCUS point is plotted first. The new CLEO2 electron plot is the 
first ``electron'' point. I also show an informal weighted average of 
these measurements including our preliminary FOCUS point.   
 \label{pdg}}
  \end{center}
\end{figure}
The CLEO Collaboration has made a new measurement of 
$\krzlnbrb{} = 0.74 \pm 0.04 \pm 0.05$ that is somewhat higher
than previous measurements and significantly higher than the previous
high precision measurement by E691 as shown in Fig. \ref{pdg}.
The new CLEO measurement can be interpreted as helping to resolve
an old problem with theory theory over-predicting the 
$\Gamma \left(\krzlndk \right)$ by a rough factor of two. 

FOCUS is in the process of making a new measurement of \krzmndk{}
using a Monte Carlo that includes the s-wave interference described above.
Our preliminary number is $\krzlnbrb{} = 0.60 \pm 0.01$ with a 
systematic error expected to be roughly twice the statistical error.
After multiplying this relative muon branching ratio by 1.05 
to compare to the electron branching ratio\cite{pdg}, our preliminary
number lies about 1.6 $\sigma$ below the new CLEO number.
\medskip

To summarize: I presented evidence for an s-wave interference with the dominant
\krzmndk{} contribution to \kpimndk{} decay.  This interference
creates a strong ($\approx 20\%$) forward-backward asymmetry in the
$\krzb{}$ decay angular distribution, but creates very minimal
distortion to the $K^- \pi^+$ mass distribution.  The dependence of the
asymmetry on the \mkpi{} suggests that it has a phase of $45^o$
near the \krzb{} pole and amplitude that is roughly 7\% of the
Breit-Wigner amplitude at the pole mass in the $H_0$ helicity
contribution. 

CLEO recently published branching ratio of \krzlndk{} relative to \kpipi{}
that was somewhat higher than the previous world average and would help
resolve a discrepancy with theoretical predictions. 
A preliminary number from FOCUS with better precision than previously
reported is 1.6 $\sigma$ lower than this CLEO number. 

We can look forward to new measurements of the \krzmndk{} form
factors, the $D_s^+ \rightarrow \phi \mu \nu / \phi \pi$ and
their form factors, studies of the \qsq{} dependence of the $D^0 \rightarrow
K^- \mu^+ \nu$ form factor, and  Cabibbo suppressed ratios
such as $D^+ \rightarrow \rho \mu \nu / \krzb \mu \nu$
and $D^0 \rightarrow \pi^- \mu^+ \nu /  K^- \mu^+ \nu$.

\bigskip
I am grateful to the FOCUS Collaboration and organizers of this
excellent conference.

\end{document}